\documentclass[longauth,letter,structabstract]{aa} 
\usepackage{graphicx}
\usepackage{txfonts}
\begin{document}
\def\HII{H\,{\sc{ii}}}
\def\mm{\,$\mu$m}
   \title{Star formation triggered by the Galactic H\,{\sc{ii}} region RCW~120}

   \subtitle{First results from the $\it{Herschel}$\thanks{Herschel is an ESA space observatory with science
instruments provided by European-led Principal Investigator
consortia and with important participation from NASA.} Space Observatory}

   \author{A. Zavagno
          \inst{1}
          \and D. Russeil \inst{1}
         \and F. Motte \inst{2} 
          \and L.D. Anderson \inst{1}
	\and L. Deharveng\inst{1}
        \and J.A. Rod\'on \inst{1} 
	\and S. Bontemps \inst{2,}\inst{3}
	\and A. Abergel \inst{4}
	\and J.-P. Baluteau \inst{1} 
	\and M. Sauvage \inst{2}
	\and P. Andr\'e\inst{2}
	\and T. Hill\inst{2}
	\and G.J. White\inst{5,}\inst{6}
          }

   \institute{ Laboratoire d'Astrophysique de Marseille (UMR 6110 CNRS \& 
Universit\'e de Provence), 38 rue F. Joliot-Curie, 13388 Marseille 
Cedex 13, France
              \email{annie.zavagno@oamp.fr}
         \and Laboratoire AIM, CEA/IRFU – CNRS/INSU – Universit\'e Paris Diderot, CEA-Saclay, F-91191 Gif-sur-Yvette Cedex, France
             \email{fmotte@cea.fr}
\and Laboratoire d’Astrophysique de Bordeaux, CNRS/INSU – Universit\'e de Bordeaux, BP 89, 33271 Floirac cedex, France 
\and Institut d'Astrophysique Spatiale, CNRS/Universit\'e Paris-Sud\,11, 91405 Orsay, France 
\and Department of Physics $\&$ Astronomy, The Open University, UK
\and Space Science Department, Rutherford Appleton Laboratory, Chilton, UK
              }  
   \date{Received March 31, 2010; accepted April 12, 2010}

 
  \abstract
   {By means of different physical mechanisms, the expansion of H\,{\sc{ii}} regions can promote the formation of new stars of all masses. RCW~120 is a nearby Galactic \HII\ region where triggered star formation occurs. This region is well-studied - there being a wealth of existing data - and is nearby. However, it is surrounded by dense regions for which far infrared data is essential to obtain an unbiased view of the star formation process and in particular to establish whether very young protostars are present.}
   {We attempt to identify all Young Stellar Objects (YSOs), especially those previously undetected at shorter wavelengths, to derive their physical properties and obtain insight into the star formation history in this region.}
   {We use $\it{Herschel}$-PACS and -SPIRE images to determine the distribution of YSOs observed in the field. We use a spectral energy distribution fitting tool to derive the YSOs physical properties. }
   {$\it{Herschel}$-PACS and -SPIRE images confirm the existence of a young source and allow us to determine its nature as a high-mass (8-10~M$_{\odot}$) Class~0 object (whose emission is dominated by a massive envelope M$_{env}$ $\simeq$ 10$^{3}$~M$_{\odot}$) towards the massive condensation~1 observed at (sub)-millimeter wavelengths. This source was not detected at 24\,$\mu$m and only barely seen in the MISPGAL 70\,$\mu$m data. Several other red sources are detected at $\it{Herschel}$ wavelengths and coincide with the peaks of the millimeter condensations. SED fitting results for the brightest $\it{Herschel}$ sources indicate that, apart from the massive Class~0 that forms in condensation 1, low mass (0.8\,-- 4~M$_{\odot}$) stars are forming around RCW~120 with ages younger than 
5$\times$10$^4$ years. This indicates that YSOs observed on the borders of RCW~120 are younger than its ionizing star, which has an age of about 2.5~Myr.}
   {$\it{Herschel}$ images allow us to detect new YSOs that are too young and embedded to be detected at shorter wavelengths (25 of the 49 $\it{Herschel}$ sources are new detections). This offers a new and more complete view of the star formation in this region. PACS and SPIRE fluxes were obtained for the brightest YSOs and allow us to strongly constrain both their spectral energy distribution and their physical properties through SED fitting. A more accurate determination of their properties allows us, for the first time, to discuss the star formation history in this region by comparing $\it{similar}$ sources at $\it{different}$ evolutionary stages. }


   \maketitle
\titlerunning{An $\it{Herschel}$ view of RCW~120}
\authorrunning{Zavagno et al.}
%
\section{Introduction}
RCW~120 (Rodgers et al.~\cite{rcw60}) is a  bubble-shaped Galactic \HII\ region, located 1.3~kpc from the Sun. 
This region was studied in detail by Zavagno et al.~(\cite{zav07}, hereafter ZA07)  
and Deharveng et al.~(\cite{DE09}, hereafter DE09). ZA07 observed the cold dust emission at 1.2~mm in this region with SIMBA at SEST and found five condensations that border the southern half of the ionized region. Three other condensations are observed away from the region (see Fig.~\ref{Fig1}). However, as is clearly observed in extinction in the $K_{\rm{S}}$-band at 2.17\,$\mu$m (see Fig.~3 in ZA07), the region is surrounded by a shell of dense material. Dark filaments 
are observed in the northern part of the region in the $\it{Spitzer}$-GLIMPSE 8\,$\mu$m image (see Fig.~3 in ZA07 and Fig.~\ref{Fig1}). ZA07 also used the 2MASS and $\it{Spitzer}$-GLIMPSE point source catalogs to study the nature and spatial distribution of the YSOs observed towards this region. They found numerous Class~I and Class~II sources towards the millimeter condensations that border the ionized region. The association of these young stellar objects (YSOs) with the RCW~120 region was then confirmed with SINFONI spectro-imaging observations (Martins et al. \cite{mar10}). ZA07 also highlighted the porous nature of the photodissociation region (their Fig.~13), which allows the stellar radiation to leak into the ambient medium. DE09 studied this region using APEX-LABOCA 870\,$\mu$m and $\it{Spitzer}$-MISPGAL observations. They refined the discussion about YSO properties by adding the 24\,$\mu$m MIPSGAL flux for the detected YSOs. They derived a mass of about 2000~M$_{\odot}$ for the shell surrounding RCW~120 and proposed the existence of a Class~0 object towards the most massive condensation (their condensation~1, M$\simeq$500~M$_{\odot}$). However, it was not possible at that time to characterize the properties of this source, which had only been detected in the MIPSGAL 70\,$\mu$m data, because no data were available at longer wavelengths. The far-IR range is indeed crucial for constraining the properties of YSOs, especially by means of spectral energy distribution (SED) fitting. 
Another point impossible to address at that time was the deeply embedded, young stellar population that may form in high-extinction regions, which had not been detected at 24\,$\mu$m. A complete census of the YSOs is needed to understand how star formation proceeds in this region. In the framework of triggered star formation it is particularly important to search for an age sequence as a function of distance from the ionization front, which may identify younger stars far from it. With its unprecedented resolution and sensitivity, the $\it{Herschel}$ Space Observatory (Pilbratt et al.~\cite{pil10}) allows us to complete this study. 

Here we present the first $\it{Herschel}$-PACS and -SPIRE images of RCW~120. We use these images to discuss the distribution and physical properties of the YSOs observed in the field. Anderson et al.~({\cite{and10}) discuss the dust properties in this region.        
\section{Observations}
RCW~120 was observed by the $\it{Herschel}$ telescope on October 9th., 2009 as part of the 
science demonstration phase with PACS (Poglitsch et al.~\cite{pog10}) for the HOBYS (Motte et al.~\cite{mot10}) Guaranteed Time Key Program (GTKP) 
and SPIRE (Griffin et al. \cite{gri10}, Swinyard et al.~\cite{swi10}) for the ISM GTKP (Abergel et al.~\cite{abe10}). Data were taken
in five wavelength bands: 100 and 160\,$\mu$m for PACS (at resolutions 10$\arcsec$\, and 14$\arcsec$), and 250, 350, and 500\,$\mu$m for SPIRE (at resolutions of 18$\arcsec$, 25$\arcsec$, and 36$\arcsec$). Two cross-scanned 30$\arcmin \times$ 30$\arcmin$ (22$\arcmin \times$ 22$\arcmin$ for SPIRE) maps at angles of 45$\degr$ and 135$\degr$ were obtained with the PACS and SPIRE instruments at a medium scan speed (30$\arcsec$/sec). We reduced the data using HIPE 2.0. The SPIRE images are the level 2 products of the SPIRE photometer pipeline. The PACS images were reduced using a script kindly provided by M. Sauvage. The most recent calibration changes for the PACS data were taken into account (see the scan map release note of February 23, 2010). Aperture photometry was performed on 49 sources in the field. All the images were convolved to match the 500\,$\mu$m image resolution. We selected a circular aperture on the sources and subtracted a local background measured within an annular aperture. 

   \begin{figure*}
   \centering
   \includegraphics[angle=0,width=180mm]{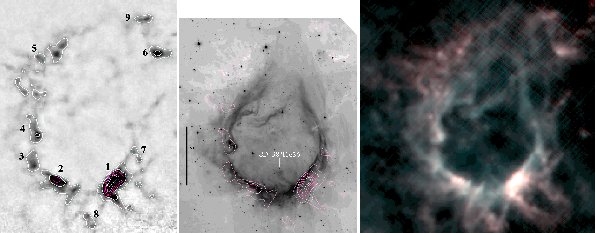}
   \caption{A multiwavelength view of RCW~120. Left: the 870\,$\mu$m APEX-LABOCA image (DE09) with 870\,$\mu$m emission contours superimposed, as in the 8\,$\mu$m image. Countour levels are from 0.15 to 2.5 Jy/beam. The condensations are identified from 1 to 9. Center: $\it{Spitzer}$-GLIMPSE 8\,$\mu$m image. The ionizing star of RCW~120, CD-38$^{\degr}$11636, is identified. Right: Colour-composite image of RCW~120 with SPIRE 250\,$\mu$m (blue) and SPIRE 500\,$\mu$m (red). The coldest parts are clearly seen in red and match very well the position of the 870\,$\mu$m condensations.}
              \label{Fig1}%
    \end{figure*}
   \begin{figure*}
   \centering
   \includegraphics[angle=0,width=180mm]{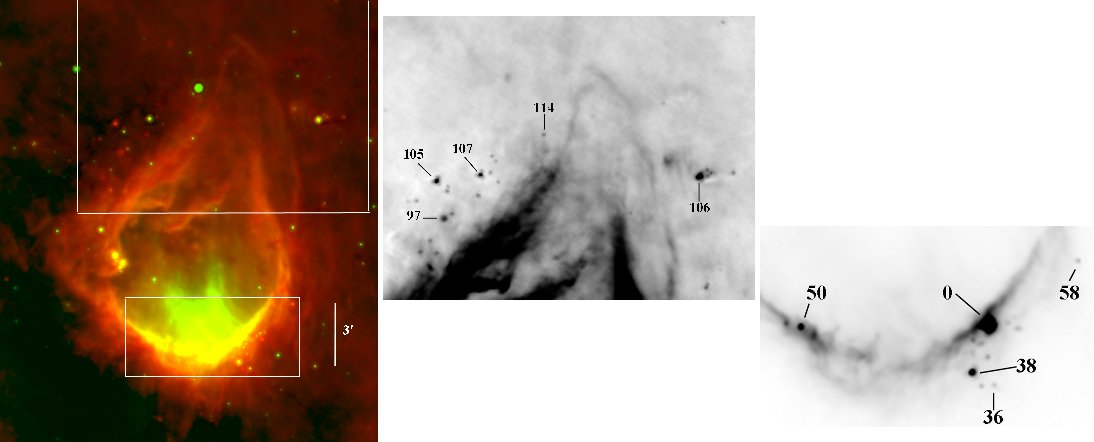}
   \caption{Left: Colour-composite image of RCW~120 of MIPSGAL 24\,$\mu$m (green) and PACS 100\,$\mu$m (red). The red sources that represent the new population of YSOs revealed by $\it{Herschel}$ are clearly seen. $\it{Herschel}$-PACS 100\,$\mu$m image: zoom on the northern part (center) and southern part (right) of RCW~120 with the brightest YSOs identified, using the numbering in DE09. Source 0 is the massive Class~0 protostellar object whose SED is given in 
Fig.~\ref{Fig3}. }
              \label{Fig2}%
    \end{figure*}

\section{$\it{Herschel}$ results}
Figure~\ref{Fig2} presents a composite image of RCW~120 at 24\,$\mu$m (green) and 100\,$\mu$m (red). The 24\,$\mu$m image shows the presence of bright sources, most of which are identified as YSOs by DE09. The PACS 100\,$\mu$m data detects bright sources corresponding to the sub-millimeter peaks, traced by the 870\,$\mu$m emission (see Fig.~\ref{Fig1}). Sources observed towards condensation 1 are shown on Fig.~\ref{Fig2}. A bright object (source 0 on Fig.~\ref{Fig2}) is clearly detected at 100 and 160\,$\mu$m with PACS and is also seen in the SPIRE data. This object is barely detected in the MIPSGAL 70\,$\mu$m data. Owing to its non-detection at shorter wavelengths, DE09 suggested that this source may be of Class~0. $\it{Herschel}$ data allow us to discuss the nature of this source by constraining its SED in the far infrared range. The SED fitting for the reddest sources detected in the field is presented in Sect.~3.1 and discussed in Sect.~4. Results of the best-fit models are given in Table~1.
Using the numbering in DE09 (see their Fig.~10 and Fig.~\ref{Fig2}), we see that sources 36 and 38 are also bright at 100\,$\mu$m. Another source is detected, in-between these two, that was not detected before. 
The PACS 100\,$\mu$m image has sufficient resolution to determine the distribution of new red objects, previously undetected at shorter wavelengths. These red sources are mainly found towards the peaks of the millimeter condensations. Source 50 seen towards the peak of condensation 2, sources 69 and 76 towards condensation 4, source 105 towards condensation 5, sources 106, 108, and 110 towards condensation 6, and source 58 towards condensation 7 are bright at 100\,$\mu$m. Eight new highly-embedded sources are detected towards condensation 5 and two new sources are detected towards condensation 8 at 100\,$\mu$m. Of the 49 sources detected and measured in PACS and SPIRE data, 25 (more than a half) were previously not (or marginally) detected at 24\,$\mu$m. Since all these new sources are probably YSOs in an early evolutionary stage, this illustrates how $\it{Herschel}$ changes our view of the star formation in this region.  

Figure~\ref{Fig1} shows a multiwavelength view of RCW~120. The $\it{Spitzer}$-GLIMPSE 8\,$\mu$m, SPIRE (250 and 500\,$\mu$m), and the APEX-LABOCA 870\,$\mu$m images are shown. The (sub)-millimeter condensations discussed in ZA07 and DE09 are identified and LABOCA 870\,$\mu$m contours are superimposed. The 8\,$\mu$m image shows the emission from polycyclic aromatic hydrocarbons that delineates the photodissociation region surrounding RCW~120. Absorption patches are clearly seen corresponding to condensations 5, 6, and 9. 

Extended emission is also seen in the PACS and SPIRE images. PACS 100 and 160\,$\mu$m emision delineate the photodissociation region. The shape of the 100\,$\mu$m extended emission outlining the PDR is surprisingly similar to that of the 8\,$\mu$m emission. This is also found for other Galactic \HII\ regions (see, for example, the case of N49, Zavagno et al.~\cite{zav10}). The dust responsible for these emissions are, however, different. The similarity observed between the two types of emission apperas to indicate that there is a temperature gradient in the form of different layers that coexist in the PDR. 

Emission at 100\,$\mu$m is observed towards the interior of the ionized region. This is probably caused by projection effect, material being emitted from the front and/or back of the bubble. The longer wavelength SPIRE images clearly trace the colder material, observed away from the ionization front (Fig.~\ref{Fig1}). YSOs are detected in there, far from the ionization front. The distribution of the extended emission can be seen in Fig.~1 of Motte et al.~(\cite{mot10}) and Anderson et al.~(\cite{and10}).    
\subsection{SED fitting of YSOs}
We performed aperture photometry on 49 sources in the field. We then used the SED-fitting tool of Robitaille et al.~(\cite{rob07}) to derive the properties of the brightest objects seen on the PACS 100\,$\mu$m image. We performed the fit from 24\,$\mu$m to 870\,$\mu$m. The parameters of the best models are given in Table~\ref{fit}. Figure~\ref{Fig3} shows the result of the fit for source~0 observed towards condensation 1.
 \begin{figure}
  \includegraphics[angle=0, width=90mm]{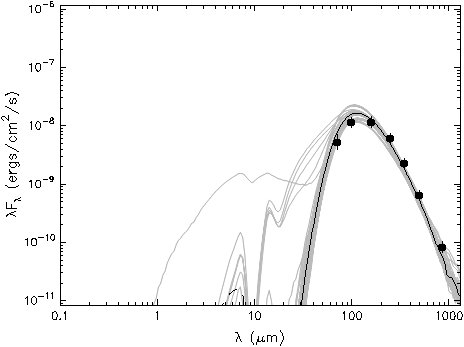}
   \caption{Spectral energy distribution obtained for source 0, the massive YSO observed towards condensation 1. The best-fit (dark solid line) function obtained using the Robitaille et al. (2007) model is shown. This is indicative of a massive (8\, -- 10~M$_{\odot}$) YSO. Its emission is dominated by its massive envelope (10$^3$M$_{\odot}$) with a high accretion rate (10$^{-3}$~M$_{\odot}$/yr). Grey lines are the models with $\chi^2-\chi^{2}_{\rm{best}}$ per data point $<3$.  }
              \label{Fig3}%
    \end{figure} 

%
\begin{table}
\caption{Parameters obtained using the SED fitting tool of Robitaille et al.~(\cite{rob07}) for the brightest PACS 
100\,$\mu$m sources }
\begin{tabular}{l r c c c}
  YSO & $M_{\rm star}$  & 
   $\dot{M}_{\rm envelope}$ & $M_{\rm envelope}$ & $L$ \\
     & ($M_{\sun}$)  & (10$^{-4}$ $M_{\sun}~yr^{-1}$) & ($M_{\sun}$)&  ($L_{\sun}$) \\
  \hline
\multicolumn{5}{l}{Condensation 1}\\
\hline
\#0  &  8--10 & 80 & 1000    & 2$\times$10$^{3}$ \\
\#36 & 1.0  & 5 & 8 & 15  \\
\#38 & 3--4  & 10 & 50--100 & 150  \\
\hline
\multicolumn{5}{l}{Condensation 2}\\
\hline
\#50 & 2--3 & 1 & 8 & 300 \\
\hline
\multicolumn{5}{l}{Condensation 4}\\
\hline
\#69 & 1.5  & 8 & 100 & 70  \\
\#76 & 0.8 & 3 & 10 & 10 \\
\hline
\multicolumn{5}{l}{Condensation 5}\\
\hline
\#105 & 1.0  & 80 & 100 & 20 \\
\hline
\multicolumn{5}{l}{Condensation 6}\\
\hline
\#106 & 1.5  & 8 & 100 & 50 \\
\hline
  \label{fit}
\end{tabular}
\end{table}
\section{Discussion}
PACS and SPIRE images confirm the existence of a young source observed towards condensation~1. These far-IR data allow us, for the first time, to determine its nature. Its emission is dominated by a massive envelope (10$^3$\,M$_{\odot}$). At 100\,$\mu$m, the source has a size (beam-corrected FWHM) of 10.9$\arcsec$, which is 0.07~pc for a distance of 1.3~kpc. We fitted its SED with a grey body to characterize its temperature. For a dust emissivity index $\beta$=2, we derived a temperature of 18~K. We computed the L$_{\rm{\lambda >350\,\mu m}}$/L$_{\rm{Bol}}$ ratio and derived a value of 0.12, characteristic of Class~0 protostars (Andr\'e et al.~\cite{and00}). All these indicators show that this source is a massive (8--10~M$_{\odot}$) Class~0 protostar. This source lies in condensation~1, elongated along the ionization front and probably formed with the material collected during the expansion of the ionized region. Source 0 is thus the first detection of a massive protostellar object formed by the collect and collapse process.       
 
Compared to the results given in DE09, the results given in Table~\ref{fit} show that the constraints on the long wavelength part of the spectrum tend to increase the value derived for the envelope accretion rate, and infer younger sources. The stellar ages span a range of between 10$^{3}$ and 5$\times$10$^{4}$ years for all the sources, but are not well constrained by the model for a given source and therefore are not included in Table~\ref{fit}. These values are well below the age of $\simeq$2.5~Myr derived for the RCW~120 ionizing star by Martins et al.~(\cite{mar10}), indicating that YSOs observed on the border of RCW~120 are indeed younger than the ionizing star.  
 
Source 105, detected farther away from the ionization front, seems to be in an earlier evolutionary stage than, for example, source 36. Both objects have stellar masses of about 1~M$_{\odot}$, but the higher accretion rate and higher envelope mass derived for source 105 point to a younger source (the possibility that source 105 is forming a higher mass star does not change the conclusion: high mass stars evolve rapidly, meaning that this source is indeed young). This is interesting, as source 105 is observed farther away from the ionization front and in a dense region that may indicate a delay in the propagation of star formation around RCW~120. However, the massive Class~0 object, source 0, is observed towards condensation~1, adjacent to the ionization front. This indicates that star formation takes more time there, even if it is close to the ionization front. 
Different physical conditions in this region (higher density, lower temperature, higher magnetic field) can decelerate the star formation process. The mass of the source also complicates a direct comparison between the different regions. 
High resolution observations using chemical and/or dynamical clocks will be performed to further investigate this point, now that we have more accurately constrained the YSO properties and can compare objects with $\it{similar}$ masses at possibly $\it{different}$ evolutionary stage. Understanding the star formation triggered by \HII\ regions are complicated by the difficulty in ascertaining the mass and evolutionary stage of YSOs observed on the borders of these regions. This problem is important when searching for age gradients in regions where triggered star formation is believed to occur.        

A question that remained unanswered by Martins et al.~(\cite{mar10}) was whether, for objects with similar spectral properties, brighter 24\,$\mu$m sources are more massive or in a later evolutionary stage. Sources 69 and 106 have similar  properties (see Table~\ref{fit}), and are both dominated by Br$\gamma$ emission in the near-IR (Martins et al.~\cite{mar10}), but source 106 is much brighter at 24\,$\mu$m than source 69. Given the new constraints derived from their SEDs, we suggest that the difference in 24\,$\mu$m emission 
is indicative of different evolutionary stages. Many other sources studied by Martins et al.~(\cite{mar10}) towards three Galactic \HII\ regions observed by $\it{Herschel}$ will be compared in a forthcoming paper. Observations using age tracers will also be peformed to confirm this hypothesis.     

\section{Conclusions}
We have presented the first images obtained with PACS and SPIRE for the Galactic \HII\ region RCW~120. These images detected an embedded population of YSOs that had not been previously detected, or had been barely detected, at wavelengths shorter than 100\,$\mu$m. The broad wavelength coverage of PACS and SPIRE allows us to constrain the SED of these sources and derive their physical properties. Our main conclusions are:  
   \begin{itemize}
      \item{We detect a massive (8--10~M$_{\odot}$) YSO towards condensation 1 and show that this source is a massive Class~0 object. This is the first detection of a massive Class~0 formed by the collect and collapse process on the borders of an \HII\ region. 
This source was barely detected previously in the MIPSGAL 70\,$\mu$m data but its mass and exact evolutionary stage were unknown. }
      \item{Our stronger constraints on the source SEDs allow us to determine more accurately the physical properties of the observed YSOs. This provides additional insight into the star formation history of this region because we can now trace (and compare) objects with similar masses but in different evolutionary stages.}   
      \item{YSOs revealed by the PACS and SPIRE are observed farther away from the ionization front, especially towards condensation 5 where dark filaments are observed in the near- and mid-IR. This suggests that triggered star formation may occur by means of the tunnelling of radiation from the ionizing star of RCW~120 into the ambient medium.}
   \end{itemize}
Six other Galactic \HII\ regions that show evidence of triggered star formation at their borders will be observed with PACS and SPIRE, in imaging and spectroscopy, as part of the HOBYS and ISM GT KPs. This will allow us to study in more detail the star formation history of these regions and place stronger constraints on the mechanism responsible for forming stars. On a global Galactic scale, the Hi-GAL Open Time Key Program (Molinari et al.~\cite{mol10}) will allow us to study the efficiency of this process in our Galaxy (Zavagno et al.~\cite{zav10}). 

\begin{acknowledgements}
   Part of this work was supported by the ANR ({\it Agence Nationale pour
  la Recherche}) project ``PROBeS'', number ANR-08-BLAN-0241 (LA). PACS
has been developed by a consortium of institutes led by MPE (Germany)
and including UVIE (Austria); KUL, CSL, IMEC (Belgium); CEA, LAM
(France); MPIA (Germany); IFSI, OAP/AOT, OAA/CAISMI, LENS, SISSA
(Italy); IAC (Spain). This development has been supported by the
funding agencies BMVIT (Austria), ESA-PRODEX (Belgium), CEA/CNES
(France), DLR (Germany), ASI (Italy), and CICT/MCT (Spain). 
SPIRE has been developed by a consortium of institutes led by
Cardiff Univ. (UK) and including Univ. Lethbridge (Canada);
NAOC (China); CEA, LAM (France); IFSI, Univ. Padua (Italy);
IAC (Spain); Stockholm Observatory (Sweden); Imperial College
London, RAL, UCL-MSSL, UKATC, Univ. Sussex (UK); Caltech, JPL,
NHSC, Univ. Colorado (USA). This development has been supported
by national funding agencies: CSA (Canada); NAOC (China); CEA,
CNES, CNRS (France); ASI (Italy); MCINN (Spain); SNSB (Sweden);
STFC (UK); and NASA (USA). 
\end{acknowledgements}


\begin{thebibliography}{}
\bibitem[2010]{abe10} Abergel, A. et al. 2010, A\&A, this volume
\bibitem[2010]{and10} Anderson, L.D., Zavagno, A., Rod\'on, J.A., et al. 2010, A\&A, this volume
\bibitem[2000]{and00} Andr\'e, P., Ward-Thompson, D., Barsony, M. 2000, in Protostars \& Planets IV, ed. V. Mannings, A. Boss, \& S. Russell (Tucson: Univ. Arizona Press), 59
\bibitem[2009]{DE09} Deharveng, L., Zavagno, A., Schuller, F., Caplan, J., Pomar\`es, M., De Breuck, C. 2009, A\&A, 496, 117 (DE09)
\bibitem[2010]{gri10} Griffin et al. 2010, this volume
\bibitem[2010]{mar10} Martins, F., Pomar\`es, M., Deharveng, L., Zavagno, A.,Bouret, J.C. 2010, A\&A, 510, 32
\bibitem[2010]{mol10} Molinari, S. et al. 2010, this volume
\bibitem[2010]{mot10} Motte, F. et al. 2010, this volume
\bibitem[2010]{pil10} Pilbratt G. et al. 2010, this volume
\bibitem[2010]{pog10} Poglitsch A. et al. 2010, this volume
\bibitem[2007]{rob07} Robitaille, T.P., Whitney, B.A., Indebetouw, R., \& Wood, K. 2007, ApJS, 169, 328 
\bibitem[2008]{rob08} Robitaille, T., Meade, M.R., Babler, B.L., Whitney, B.A., Johnston, K.G., et al. 2008, AJ, 136, 2413
\bibitem[1960]{rcw60} Rodgers, A.W., Campbell, C.T., Whiteoak, J.B. 1960, MNRAS, 121, 103
\bibitem[2010]{swi10} Swinyard, B., Ade, P., Baluteau, J.-P. et al. 2010, A\&A, this volume
\bibitem[2007]{zav07} Zavagno, A., Pomar\`es, M., Deharveng, L., Hosokawa, T., Russeil, D., Caplan, J. 2007, A\&A, 472, 835
\bibitem[2010]{zav10} Zavagno, A. et al. 2010, A\&A this volume
\end{thebibliography}
\end{document}